\documentclass{emulateapj}

\usepackage{url}
\usepackage{natbib}
\newcommand{\beq}{\begin{equation}}
\newcommand{\eeq}{\end{equation}}
\newcommand{\bdi}{\begin{displaymath}}
\newcommand{\edi}{\end{displaymath}}
\newcommand{\MSX}{\emph{MSX}}
\newcommand{\IRAS}{\emph{IRAS}}
\newcommand{\MIPS}{MIPS}
\newcommand{\IRAC}{IRAC}
\newcommand{\SCUBA}{SCUBA} %{\emph{SCUBA}}

\newcommand{\kms}{km\,s$^{-1}$}

\newcommand{\HII}{\ion{H}{2}}
\newcommand{\HI}{\ion{H}{1}}
\newcommand{\msol}{M$_{\odot}$}
\newcommand{\lsol}{L$_{\odot}$}
\newcommand{\LM}{$L$-$M$}
\slugcomment{Submitted to the Astrophysical Journal}

\begin{document}

%\shorttitle{L-R deconvolution for BLAST images}
\shorttitle{K3-50 and IC~5146 in deconvolved BLAST images}
\shortauthors{Roy, A.~et al.}

%% This is the end of the preamble.  Indicate the beginning of the
%% paper itself with \begin{document}.


\title{Deconvolution of images from BLAST 2005: Insight into the K3-50
  and IC~5146 star-forming regions}

\author{Arabindo~Roy,\altaffilmark{1,\dag}
        Peter~A.~R.~Ade,\altaffilmark{2}
        James~J.~Bock,\altaffilmark{3,4}
        Christopher~M.~Brunt,\altaffilmark{5}
        Edward~L.~Chapin,\altaffilmark{6}
        Mark~J.~Devlin,\altaffilmark{7}
        Simon~R.~Dicker,\altaffilmark{7}
        Kevin~France,\altaffilmark{8}
        Andrew~G.~Gibb,\altaffilmark{6}
        Matthew~Griffin,\altaffilmark{2}
        Joshua~O.~Gundersen,\altaffilmark{9}
        Mark~Halpern,\altaffilmark{6}
        Peter~C.~Hargrave,\altaffilmark{2}
        David~H.~Hughes,\altaffilmark{10}
        Jeff~Klein,\altaffilmark{7}
        Gaelen~Marsden,\altaffilmark{6}
        Peter~G.~Martin,\altaffilmark{11}
        Philip~Mauskopf,\altaffilmark{2}
        Calvin~B.~Netterfield,\altaffilmark{1,12}
        Luca~Olmi,\altaffilmark{13,14}
        Guillaume~Patanchon,\altaffilmark{15}
        Marie~Rex,\altaffilmark{7}
        Douglas~Scott,\altaffilmark{6}
        Christopher~Semisch,\altaffilmark{7}
        Matthew~D.~P.~Truch,\altaffilmark{7}
        Carole~Tucker,\altaffilmark{2}
        Gregory~S.~Tucker,\altaffilmark{16}
        Marco~P.~Viero,\altaffilmark{1,4}
        Donald~V.~Wiebe\altaffilmark{6}}

\altaffiltext{1}{Department of Astronomy \& Astrophysics, University of
  Toronto, 50 St. George Street, Toronto, ON  M5S~3H4, Canada}

\altaffiltext{2}{Department of Physics \& Astronomy, Cardiff University,
  5 The Parade, Cardiff, CF24~3AA, UK}

\altaffiltext{3}{Jet Propulsion Laboratory, Pasadena, CA 91109-8099}

\altaffiltext{4}{Observational Cosmology, MS 59-33, California Institute
  of Technology, Pasadena, CA 91125}

\altaffiltext{5}{School of Physics, University of Exeter, Stocker Road,
  Exeter EX4 4QL}

\altaffiltext{6}{Department of Physics \& Astronomy, University of
  British Columbia, 6224 Agricultural Road, Vancouver, BC
  V6T~1Z1,Canada}

\altaffiltext{7}{Department of Physics and Astronomy, University of
  Pennsylvania, 209 South 33rd Street, Philadelphia, PA 19104}

\altaffiltext{8}{Center for Astrophysics and Space Astronomy, University
  of Colorado, Boulder CO, 8030}

\altaffiltext{9}{Department of Physics, University of Miami, 1320 Campo
  Sano Drive, Carol Gables, FL 33146}

\altaffiltext{10}{Instituto Nacional de Astrof{\'i}sica {\'O}ptica y
  Electr{\'o}nica (INAOE), Aptdo. Postal 51 y 72000 Puebla, Mexico}

\altaffiltext{11}{Canadian Institute for Theoretical Astrophysics,
  University of Toronto, 60 St. George Street, Toronto, ON M5S~3H8,
  Canada}

\altaffiltext{12}{Department of Physics, University of Toronto, 60
  St. George Street, Toronto, ON M5S~1A7, Canada}

\altaffiltext{13} {University of Puerto Rico, Rio Piedras Campus,
  Physics Dept., Box 23343, UPR station, San Juan, Puerto Rico}

\altaffiltext{14}{Istituto di Radioastronomia, Largo E. Fermi 5,
  I-50125, Firenze, Italy}

\altaffiltext{15}{Laboratoire APC, 10, rue Alice Domon et L{\'e}onie
  Duquet 75205 Paris, France}

\altaffiltext{16}{Department of Physics, Brown University, 182 Hope
  Street, Providence, RI 02912}

\altaffiltext{\dag}{\url{aroy@cita.utoronto.ca}}

\begin{abstract}

We present an implementation of the iterative flux-conserving
Lucy-Richardson (L-R) deconvolution method of image restoration for maps
produced by the Balloon-borne Large Aperture Submillimeter Telescope
(BLAST).  Compared to the direct Fourier transform method of
deconvolution, the L-R operation restores images with better-controlled
background noise and increases source detectability.  Intermediate
iterated images are useful for studying extended diffuse structures,
while the later iterations truly enhance point sources to near the
designed diffraction limit of the telescope. The L-R method of
deconvolution is efficient in resolving compact sources in crowded
regions while simultaneously conserving their respective flux densities.
We have analyzed its performance and convergence extensively through
simulations and cross-correlations of the deconvolved images with
available high-resolution maps.  We present new science results from two
BLAST surveys, in the Galactic regions K3-50 and IC~5146, further
demonstrating the benefits of performing this deconvolution.

We have resolved three clumps within a radius of 4\farcm5 inside the
star-forming molecular cloud containing K3-50. Combining the
well-resolved dust emission map with available multi-wavelength data, we
have constrained the Spectral Energy Distributions (SEDs) of five clumps
to obtain masses ($M$), bolometric luminosities ($L$), and dust
temperatures ($T$). The \LM\ diagram has been used as a diagnostic tool
to estimate the evolutionary stages of the clumps.  There are close
relationships between dust continuum emission and both 21-cm radio
continuum and $^{12}$CO molecular line emission.

The restored extended large scale structures in the Northern Streamer of
IC~5146 have a strong spatial correlation with both \SCUBA\ and high
resolution extinction images. A dust temperature of 12~K has been
obtained for the central filament. We report physical properties of ten
compact sources, including six associated protostars, by fitting SEDs to
multi-wavelength data. All of these compact sources are still quite cold
(typical temperature below $\sim$ 16~K) and are above the critical
Bonner-Ebert mass. They have associated low-power Young Stellar Objects
(YSOs).  Further evidence for starless clumps has also been found in the
IC~5146 region.
\end{abstract}


\section{Introduction}     

The BLAST flight of 2005 (BLAST05; \citealp{pascale2008, chapin2008})
conducted both targeted and unbiased surveys in some of the important
star forming regions of the Galactic Plane. These were conducted
simultaneously at the submillimeter wavelengths of 250, 350, and
500~\micron. BLAST was a precursor to \emph{Herschel} observations with
SPIRE \citep{griffin2010}.
One of the science goals of BLAST surveys was to study the early stages
of massive star formation and identify their precursors in order to
describe a complete sequence of evolution. In order to realize this
ambition, a deep survey complemented by high spatial resolution is
required.
With a 2-m telescope, BLAST05 optics were designed to produce dust
emission maps at a diffraction-limited resolution of full width at half
maximum (FWHM) 40, 58, and 75\arcsec\ at 250, 350, and 500~\micron,
respectively \citep{pascale2008}. Unfortunately,
due to some uncharacterized optical problem, the BLAST05 Point Spread
Function (PSF) was anomalous, producing images at a resolution of about
3\farcm3 \citep{truch2008, roy2010}.

Most of the resulting maps contained clear imprints of the corrupted PSF
(hexagonal geometry), indicating the presence of compact sources with
angular sizes comparable to the diffraction limit. Measurements of the
flux density of a few isolated bright sources can be carried out by
aperture photometry or by fitting the PSF.  However, because of the
large beam area of the corrupted PSF, the sensitivity to compact point
sources in the presence of cirrus noise suffers, and the detection of
crowded sources is challenging. This leads to an underestimate of the
source density and produces an incomplete source list.

In order to improve the image resolution to near the diffraction limit,
a direct Fourier transform method of deconvolution was performed in
analyzing the Vulpecula region \citep{chapin2008}.
In this method of deconvolution, compact sources in the processed map
were enhanced to 1\farcm3, 1\farcm6, and 1\farcm8 at 250, 350, and
500~\micron, respectively. However, the deconvolved maps were
contaminated by ripples which propagated across the entire map.  The
generation of ripples is an intrinsic consequence of the direct Fourier
inversion method when the PSF varies across the field and is not
perfectly known.  The amplitude of the ripples is particularly high near
bright sources. Strong interference patterns are also observed in the
vicinity of multiple bright sources.  This makes fainter compact sources
hard to detect; furthermore, artificial spatial features are introduced
in diffuse structures.
 
To get the best science return from the BLAST maps, we consider
implementation of the L-R method of deconvolution
\citep{richardson1972,lucy1974}.
The L-R deconvolution method is an iterative way of estimating the sky
brightness distribution, conserving flux at every successive step.
Fortunately, maps from the BLAST 2005 flight have a high signal-to-noise
ratio and are well sampled with 15\arcsec\ pixels, which is particularly
important for the performance of L-R deconvolution.

Implementation of L-R deconvolution on the BLAST maps, for example Cas~A
\citep{sibthorpe2010}, Aquila \citep{rivera2010}, and Cygnus X
\citep{roycyg2010}, has remarkably enhanced the compactness of point
sources to about 1\arcmin\ effective resolution. The other benefits of
the L-R scheme are: 1) preservation of spatially diffuse structures; 2)
suppression of background noise; and 3) resolving blended sources. As an
example, in the Cygnus~X region the L-R map led to a source catalog
containing 184 sources, significantly greater than the 130 produced
using the direct Fourier inversion method.
 
A basic goal of this paper is to demonstrate the performance of the L-R
algorithm in deconvolving BLAST05 maps. We do this via simulations and
using ``ground truth,'' i.e., comparing the details in BLAST deconvolved
images with available high-resolution maps, thereby establishing the
reliability of detected structures, both compact
(\S~\ref{sec:lr-on-srcs}) and more diffuse
(\S~\ref{sec:lr-on-diffuse}). In those subsections we focus primarily on
the operational consequences of deconvolution for identifying compact
sources and defining diffuse structures.
We subsequently present new science results made possible by this
deconvolution, for the star-forming regions K3-50 (\S~\ref{sec:k350})
and IC~5146 (\S~\ref{sec:ic5146}).  We identify protostellar clumps and
characterize their physical properties such as temperature ($T$), mass
($M$), and luminosity ($L$), by constraining spectral energy
distributions (SEDs) using multi-wavelength data.  We have also
investigated the properties in the ``Northern Streamer,'' a cold
molecular ridge in the IC~5146 field.

\section{ Overview of Deconvolution}

Deconvolution is a common technique used by the astronomical community
for enhancing image quality. In the imaging process, various factors are
responsible for the distortion of the observed sky map.  In the
traditional way, the data recorded represent the sky convolved with a
point spread function (PSF). The PSF shape depends not only on the
mirror and aperture (window) of the telescope, but also often on the
atmosphere.  Even in an ideal instrumental setup, images are blurred due
to secondary peaks of Rayleigh diffraction.  In modern astronomy,
spatial resolution plays a pivotal role in deciphering morphology and
obtaining accurate photometry and hence it is important to try to
de-blur images.
 
\citet{meinel} has shown a wide variety of both linear and nonlinear
restoration algorithms, based on maximum likelihood and recursive
improvement of images in successive iterations, taking account of the
Gaussian and Poisson noise processes.  Moreover, the additional \emph{a
  priori} information about the positivity of the true image helps to
improve resolution.  One of the most frequently-used methods is the
above-mentioned L-R deconvolution scheme, which conserves flux. A key
property of this algorithm is that it converges toward the most likely
solution of the PSF and image intensity distribution.  In the
literature, there exist several other techniques for image restoration.
A detailed description of the Maximum Entropy Method (MEM) of
deconvolution is given in \citet{narayan} and \citet{bryan}. A Maximum
Correlation Method (MCM), which maximizes the correlation between
adjacent pixels, and simultaneously improves image resolution has been
been prescribed by \citet{aumann1990} for improving \IRAS\ survey
maps. This algorithm is also an extension of the L-R solution as
described in \cite{meinel} and \citet{hires}.  The latter authors
adopted the MCM technique for deconvolution of \emph{Spitzer} images,
improving the resolution by a factor of about three.

These techniques generally require that the PSF be previously determined
to high accuracy.  When the PSF is only poorly known, an iterative blind
deconvolution technique can be used for image restoration, where both
the data and the PSF are estimated simultaneously in the successive
steps \citep{fish,tsumuraya}.

\subsection{Two Deconvolution Algorithms}\label{lralgo}
 
In the classical approach to the imaging process, the data image $D$ is
expressed by
\begin{equation}
D(x,y)=I(x,y) \ast P(x,y) + N(x,y),
\label{eq:img}
\end{equation}
where $P(x,y)$ is the PSF of the telescope, $I(x,y)$ is the (unknown)
unblurred sky image, the symbol $\ast$ denotes the convolution operator,
and $N(x,y)$ is the noise added to the data while imaging. Recursive
formulae can be obtained from the maximum likelihood solution for two
different classes of noise distribution, namely the Poisson and the
Gaussian noise processes \citep{meinel,lridl}. From equations~(31) and
(64) of \citet{meinel} we write directly the solution for the Poisson
noise case in the form \citep{lridl}:
\begin{equation}
I_{\rm new} = I_{\rm{old}}\left[\left(\frac{D}{I_{\rm{old}}\ast P
  }\right) \ast \tilde P\right]^q,
\label{eq:poisson}
\end{equation}
and that for Gaussian noise as:
\begin{equation}
I_{\rm{new}}=I_{\rm{old}}+\left(\left[D-\left(I_{\rm{old}}\ast
  P\right)\right]\ast\tilde P\right)^q.
\label{eq:gaussian}
\end{equation}
Here $ \tilde P$ is the reflected PSF, $ \tilde P(x)$ $=$ $P(-x)$.
For $q~=~1$, equation~(\ref{eq:poisson}) reduces to the L-R algorithm
obtained independently from Bayesian statistics
\citep{richardson1972,lucy1974}.  The Poisson noise process, by its
nature, excludes any negative solutions whereas the Gaussian noise
process allows negative solutions, subject to the choice of $q$ in
equation~(\ref{eq:gaussian}).

Ideally, after an infinite number of iterations, the output map should
converge to the maximum-likelihood solution. However, in practice, after
a finite number of iterations in which a close to maximum-likelihood
solution has been achieved, the smoothness of the map starts
deteriorating \citep{hires}. There are no definite and generalized
stopping criteria for avoiding this \citep{prasad}.  \citet{lucy1974}
recommended a stopping criterion based on a goodness-of-fit test.
However, subsequently \citet{lucy1992} noted that a higher number of
iterations might be needed for images with a large number of pixels, and
he discouraged the use of the former method.

\subsection{Implementation on BLAST maps}\label{lrblast}

% Making the maps
BLAST05 raster-scanned targeted regions in the Galactic Plane, having
different areas ranging from 3~deg$^2$ to 10~deg$^2$.  From the
time-ordered data, maps were produced using the optimal map-maker
SANEPIC \citep{pat08}.  The combination of high scan speed and low $1/f$
knee, cross-linking where available, together with common-mode removal
in SANEPIC, produces maps retaining diffuse low spatial frequency
emission. However, preprocessing of the time-ordered data to remove very
low frequency drifts, plus using a low pass cutoff, makes the SANEPIC
map average zero -- the DC level is not known.
% adding the offset
The iterative L-R deconvolution algorithm for Poisson noise
(eq.~[\ref{eq:poisson}]) converges only for true positive images. In
order to satisfy the positivity of the initial data image $D(x,y)$, we
add a constant level to the whole map (a deconvolution operation on a
flat map preserves the initial map in its output).  The value of the
constant is estimated by pixel-pixel correlation of the BLAST images
(smoothed to 4\arcmin) with the corresponding 100~\micron\ IRIS image,
\IRAS\ data reprocessed by \citet{mairis}.  We carried out
deconvolutions with different values of the constant, finding the
solution to be robust.  After the deconvolution operation we subtracted
the constant from the resulting images.

% Preparation of the maps: Padding and apodizing
Special care has been taken to deal with the edges of the scanned
region, padding the external area at a value equal to the average around
the edge of the map.  The convolution operations in the iterative
updates were carried out by fast Fourier transform (FFT), and so to
maintain a periodic boundary condition and smoothness we apodized the
entire map edge with a sine function.

% preparation of the PSF
The maps can be produced at various pixelizations, usually 15\arcsec,
but also 10\arcsec\ or 20\arcsec.  This is sufficient to sample the
degraded PSF and to sample and recover diffraction-limited information
even in the 250~\micron\ channel.  Our experiments showed that the
deconvolution is robust to different pixelizations in this range.
Note that the corrupted PSF is not azimuthally symmetric.  Therefore, at
each passband a synthetic PSF appropriate to the particular scan pattern
and coverage of a field is constructed from the well-sampled
(10\arcsec\ grid) telescope-frame PSFs of each bolometer
\citep{chapin2008}.  In order to avoid artefacts at large angular scales
we have apodized the boundaries of the PSFs with a cos$^2$ taper from a
radius of 4\farcm5 to 5\farcm7.

% the algorithms
For the actual processing we have used the IDL-implemented
`Max\textunderscore likelihood.pro' routine by \cite{lridl}, which
includes both methods of iterative update, appropriate to Gaussian or
Poisson noise.
% noise
Probably neither is an accurate noise model for the BLAST Galactic maps,
which incorporate a fluctuating cirrus background, bright cirrus
structures, and strong compact sources.  Empirically, the largest errors
in the maps are near the bright sources, this arising from small
pointing errors, the asymmetric PSF, and the particulars of the
map-making process itself.  Furthermore, in the reconstruction step
implicit in the deconvolution (eq.~[\ref{eq:img}]), whenever the
synthetic PSF is not a perfect representation there are larger errors in
predicting $D$ where $I$ is brighter.  Thus of the two, the Poisson
algorithm is expected to work better because it gives less weight where
there are strong point sources, i.e., where there are larger errors in
the maps.  Although the noise is not actually Poisson and the solution
might not be optimal, it is always consistent with the data through
equation~(\ref{eq:img}).  Note that a similar situation arises in the
HIRES processing of \IRAS\ data, where the noise model is likely a
complicated hybrid \citep{aumann1990}.
We have performed experiments with both noise models and, as
anticipated, the Poisson alternative was found to perform better in
enhancing the structure in the image.

\begin{figure}[t]
\includegraphics[width=\linewidth]{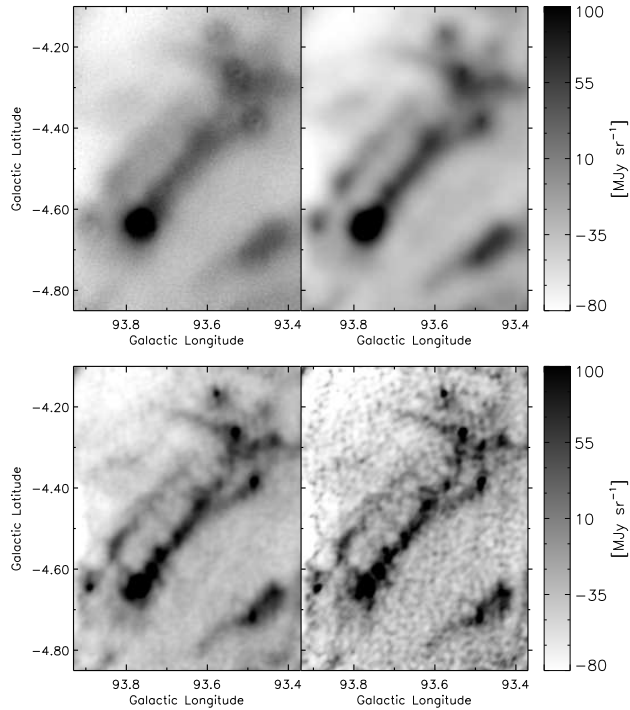}
\caption{Example of BLAST image deconvolution by the L-R algorithm
  (Poisson noise), at 250~\micron\ for the IC~5146 region.  Upper left:
  unprocessed image where compact sources have the characteristic
  imprint of the corrupted BLAST PSF. Upper right: deconvolved image
  after 8 iterations. Lower left and lower right: after 64 and 256
  interations, respectively. The typical size of the compact sources in
  the final map is about 1\arcmin.}
\label{fig:ic5146}
\end{figure}

% tests involving restoring beams
% describe more (done)
As shown in Figure~\ref{fig:ic5146}, L-R deconvolution progressively
enhances an image, but at the expense of background noise amplification
at high frequencies after a large number of iterations.  This noise can
be suppressed by convolving the deconvolved map with a restoring beam
$G$ of FWHM close to the diffraction limit.  Mathematically this step of
restoring the image might be achieved in two distinct ways: 1) convolve
each side of equation~(\ref{eq:img}) with $G$, and treat $I\ast G$ as
the unknown sky map to be found by the iterative solution of this new
equation; or alternatively 2) as the effective PSF use
$\widehat{(\widehat{P}/\widehat{G})}(x,y)$, where $\widehat{X}$ denotes
the Fourier transform of $X$.  We have compared the results for these
two alternatives with the results of the standard deconvolution
subsequently smoothed by the restoring beam and have found that L-R
deconvolution gives robust results for all three cases.  The images
shown in this paper have not been restored.

% overall assessment
The reason for this is that despite the complicated and asymmetric PSF
and the noise in the maps, application of L-R deconvolution removed the
worst effects of the corrupted beam, thus improving significantly the
definition of both compact sources and diffuse structures. Compared to
the results of the direct inversion, the background obtained is much
smoother, although ringing around bright sources has not been completely
removed. Such ringing is not seen in Figure~\ref{fig:ic5146}, where
there are no very bright sources, but can be seen in the maps of Aquila
\citep{rivera2010}, Cygnus X \citep{roycyg2010}, and K3-50 (below), each
of which contain significantly brighter sources.  This same effect is
shown in the simulations below.

% ringing
This ringing is the principal artefact in the deconvolved images, as was
also found to be the case in using the HIRES MCM algorithm to produce
the Infrared Galaxy Atlas (IGA) from \IRAS\ scans of the Galactic plane
\citep{cao1997}.  \cite{hires} reduced the ringing artefacts in
deconvolutions of \emph{Spitzer} data by first subtracting the
background around the targeted compact sources.  However, the large maps
from the BLAST surveys are more complex, having large-scale structures
of different brightness on which stronger sources are superimposed.
Therefore, it is not practical to reduce the background everywhere to be
close to zero.

These artefacts are on the scale of the corrupted PSF, and consist
initially of a ring-shaped depression with an outer enhanced ridge. In
successive iterative steps the artefact evolves, developing a finer
series of depressions and ridges of smaller amplitude spreading from the
source out into the rest of the image. The average brightness enclosed
within the ringing pattern is very close to the local sky background.
The intensity ratio of the brightest, smallest ring to the deconvolved
central peak depends somewhat on the local background, but is typically
no larger than 0.1\%.  Nevertheless, near very strong sources it is
quite obvious, adding uncertainty in the estimation of the sky
background level near bright (often crowded) sources and limiting the
detectability of nearby faint sources.

\section{Tests and Ground Truth}\label{truth}

\subsection{Simulations}

We have performed simulations to check the robustness of the
L-R deconvolution scheme.  
Fake sources were constructed by convolving normalized narrow Gaussians
with the BLAST PSF and then multiplying them with random flux
densities. Gaussians with intrinsic FWHMs ranging from 0.5\arcmin\ to
2\arcmin\ were tested. These simulated sources were inserted at different
locations in the map such that the sources were well separated from each
other.  During the BLAST map-making process, a variance map is obtained from 
a combination of the noise in the time stream data and the map coverage by 
the bolometers in the array.  We used this variance estimate to add noise to 
the simulated map, but in these tests we did not take into account the 
background structures and fluctuations from the diffuse cirrus.

\begin{figure}[t]
\includegraphics[width=\linewidth]{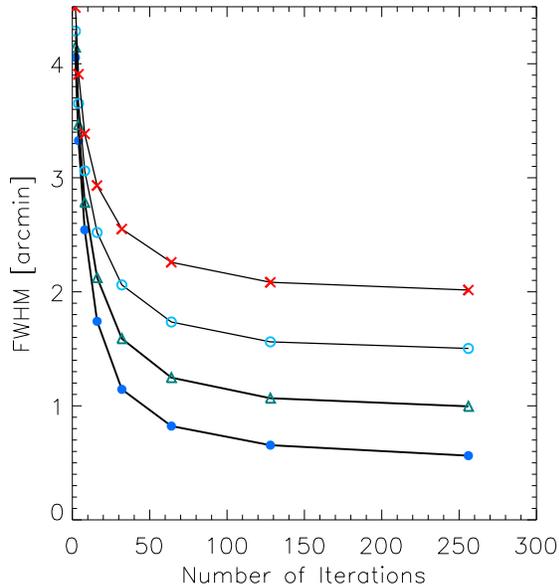}
\caption{Measured FWHM of simulated sources at successive iterations of
  L-R deconvolution. Filled circles, triangles, open circles, and
  crosses represent sources of intrinsic size 0\farcm5, 1\farcm0,
  1\farcm5, and 2\farcm0, respectively.}
\label{fig:fwhm}
\end{figure}

\begin{figure}[b]
\includegraphics[width=\linewidth]{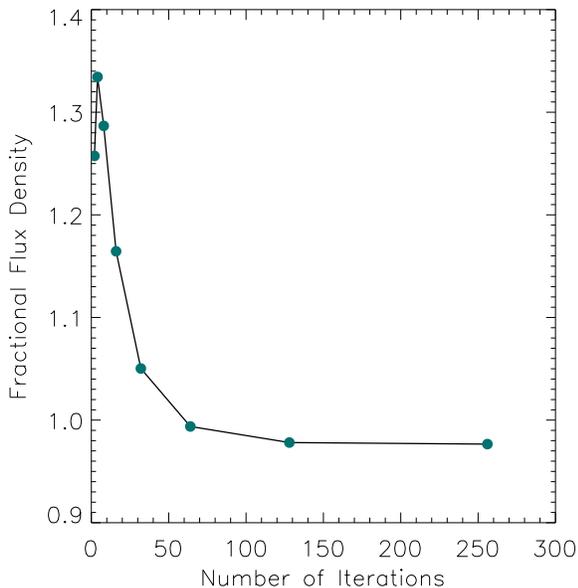}
\caption{Measured fractional flux density averaged for compact sources
  with various FWHMs, at successive iterations of L-R deconvolution.}
\label{fig:flux}
\end{figure}

We then deconvolved the simulated map using the PSF and observed the
convergence of the measured FWHMs of the sources as a function of
iteration, as shown in Figure~\ref{fig:fwhm}.  In these tests we were
able to recover the intrinsic size of the sources, convergence occurring
after about 64 iterations. In Figure~\ref{fig:flux} we have plotted the
average fractional flux densities recovered at different iterations of
the L-R deconvolution.  For isolated sources with various FWHMs about
98\% of the input flux density is recovered.  It is possible that the
remaining flux density is lost due to the ringing artefact near the
source, for which a Gaussian profile used in the fitting is not an
accurate model.

\begin{figure}[t]
\includegraphics[width=\linewidth]{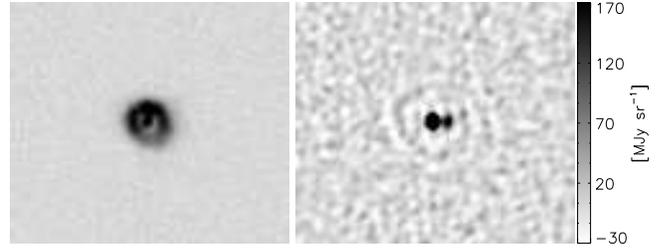}
\caption{Left panel: two simulated sources of relative flux density 10
  placed 1\farcm5 apart. Right panel: deconvolved map showing the two
  resolved sources. The brightness scale has been chosen to highlight the
  faint ring artefact.}
\label{fig:doublefig}
\end{figure}

Figure~\ref{fig:doublefig} shows the performance of L-R deconvolution in
a simulation in which two compact sources with relative flux density 10
are placed only 1\farcm5 apart, well within the corrupted PSF. In the
blurred map it is hard to discern a faint second source hidden within
the brighter source.  However, L-R deconvolution makes it clear.
Fitting a double Gaussian to the deconvolved map, we have retrieved the
initial flux densities and sizes within 96\% and 99\%, respectively.
The positions of the recovered sources are accurate within $\sim$ 1\arcsec.

\subsection{L-R Performance on Compact Sources}\label{sec:lr-on-srcs}

Application of the L-R scheme to BLAST05 maps of the Aquila and Cygnus~X
regions has already shown impressive improvement in resolving confused
sources in crowded regions \citep{rivera2010,roycyg2010}, thereby
enabling the preparation of deeper catalogs. From surveys at both longer
and shorter wavelengths there was abundant ground truth for the fainter
sources recovered.
To demonstrate the L-R performance on a field of compact sources we have
selected the region near K3-50, which is a young star-forming site
containing a group of \HII\ regions. We make use of images from
\MSX\ \citep{mill1994,ega98} at 8~\micron, 
the IGA at 60 and 100~\micron, and
the CGPS 21-cm radio continuum survey \citep{tay2003}.  We also make use
of spectral line cubes of $^{12}$CO from the Five College Radio
Astronomy Observatory
(FCRAO)\footnote{http://www.astro.umass.edu/$\sim$fcrao/} and of
\HI\ from the CGPS \citep{tay2003}.  The science results for K3-50 are
deferred to \S~\ref{sec:k350}.

\begin{figure*}[t]
\includegraphics[width=\linewidth]{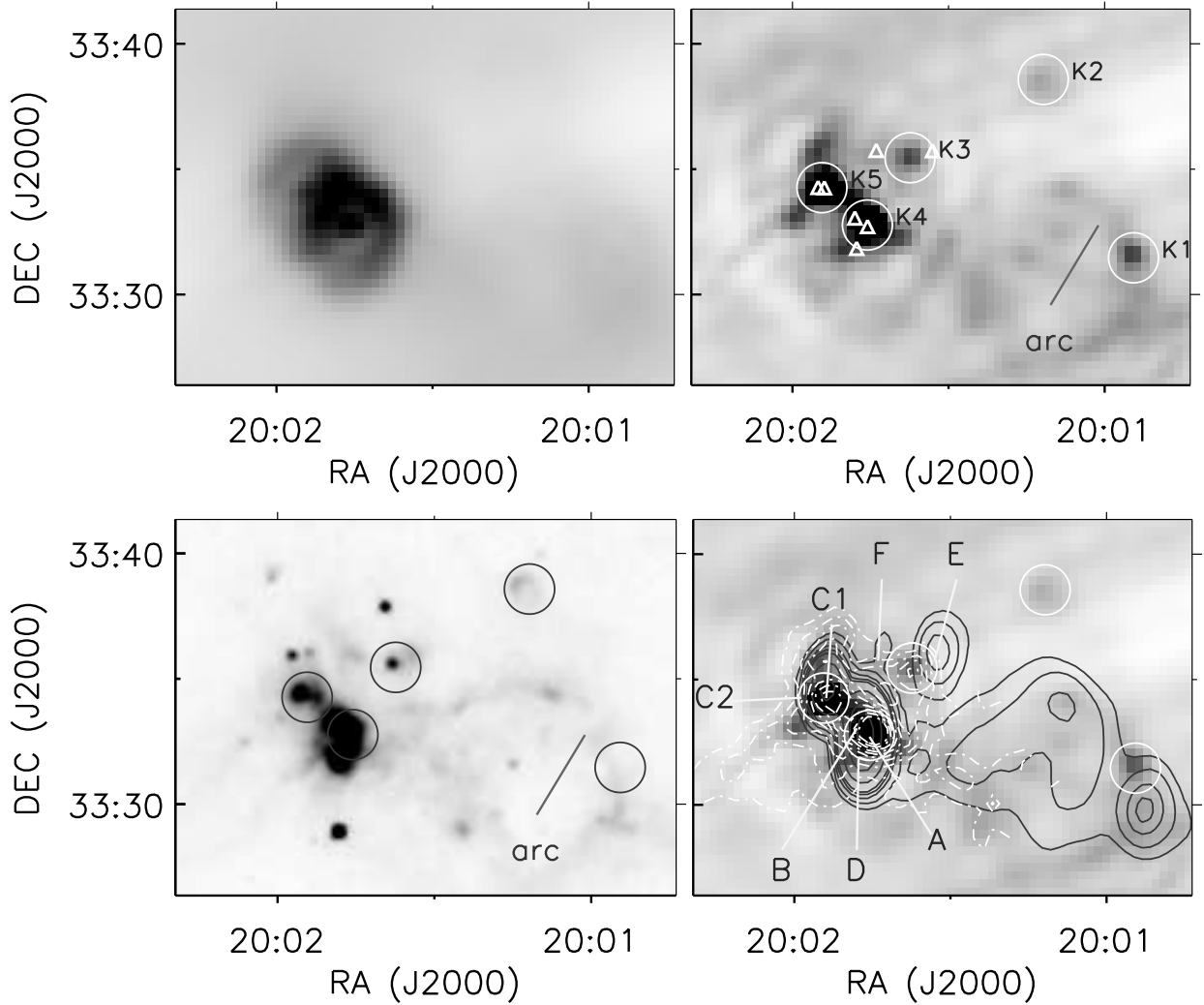}
\caption{Upper left: raw BLAST05 map of K3-50 region at 250~\micron,
  showing that crowded sources remain unresolved.
Upper right: same region after L-R deconvolution. BLAST compact sources
are shown by the circles. The recovered FWHM of K4 (K3-50A) is
36\arcsec. Triangles show the positions of radio sources from
\cite{samal2010}, with labels A to F (K3-50A to K3-50F) in the lower
right panel.
Lower left: 8~\micron\ \MSX\ image. The arc-like feature is also seen in
the BLAST image.
Lower right: solid contours from 21-cm radio continuum (CGPS) overlaid
on the deconvolved 250~\micron\ BLAST image. The dash-dot contours are
the $^{12}$CO line emission (FCRAO) integrated over $-$30 to $-$16
km~s$^{-1}$.  Note that the CO map coverage does not extend to K1 and
K2.}
\label{fig:w58}
\end{figure*}

Figure~\ref{fig:w58} shows details of the K3-50 region.  The upper left
panel is the original SANEPIC 250~\micron\ BLAST map, clearly showing
imprints of overlapping corrupted PSFs.  The upper right panel shows how
the L-R deconvolution operation has improved the resolution, enabling
the detection of faint sources which were otherwise hidden. The compact
sources have an angular resolution of $\sim$ 40\arcsec, which
corresponds to a spatial extent of about 1.7 pc at a distance of 8.7 kpc
\citep{peeters2002}.  The overlaid triangles identify the radio source
components obtained by \cite{samal2010} from 1280 MHz observations,
showing that there are multiple sources within the two bright BLAST
sources, still unresolved at the improved BLAST resolution. Their names
are indicated in the lower right panel, which also shows radio emission
and CO contours.
The lower left panel shows the same region imaged by \MSX\ (8~\micron),
also at relatively high resolution. Again, there is good correspondence
with BLAST.

K3, called IRASB by \cite{samal2010}, coincides with the point source
IRAS~19597+3327A.  At its position there is a CO peak at
$-23$~km~s$^{-1}$, the same velocity as for the much brighter K4 and K5
sources, and in the CGPS \HI\ emission there is an absorption reversal at
the same velocity.  This indicates that K3 is related to this complex
and is not a galaxy, as it has been alternatively classified
(2MASSX~J20013735+3335282).
K2 is an asymmetric ring at 8~\micron\ and has no radio counterpart.
K1 is on an arc of emission seen in the BLAST and \MSX\ 8~\micron\ maps,
apparently related to the bubble of extended \HII\ emission and its
confinement.  However, there is no prominent counterpart in either
\MSX\ or \IRAS\ images. There is no 21-cm radio continuum counterpart
either, all indicating that the source is cold and unevolved (see
below).  There is no confirming information from CO, since
unfortunately, K1 (and also K2) is outside the area covered by that
emission-line survey; the low resolution integrated CO map of
\cite{dame2001} hints at an extension in this direction.

\subsection{L-R Performance on Diffuse Structures} \label{sec:lr-on-diffuse}

\begin{figure}
\includegraphics[width=\linewidth]{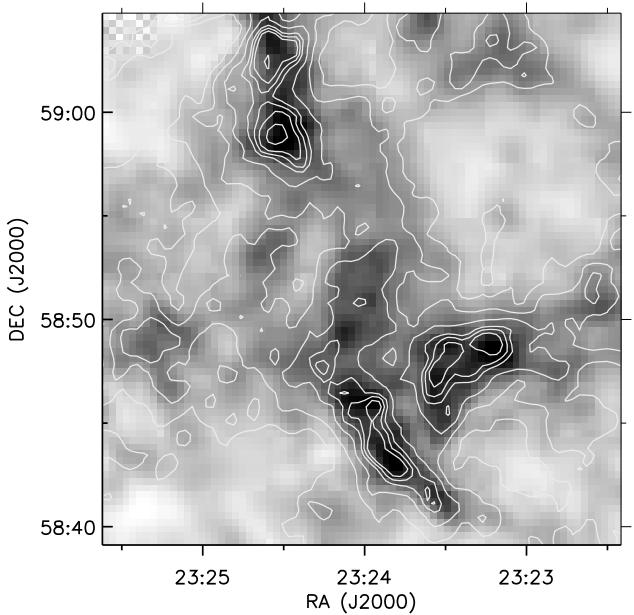}
\caption{SPIRE 250~\micron\ image of the Cas~A field convolved with
a  40\arcsec\ Gaussian and reprojected on the 20\arcsec\ BLAST grid. 
The contour overlay is from the corresponding BLAST deconvolved map, 
after 32 iterations. After deconvolution, the BLAST image faithfully
reveals the morphology of the diffuse structures.}
\label{fig:casa}
\end{figure}

In this section we discuss the effectiveness of L-R deconvolution for
recovering the morphology of extended structures.
% CasA, then new fields
BLAST05 mapped a diffuse field toward the Cas~A supernova remnant
\citep{sibthorpe2010}.  The map was scanned in only one principal
direction, but nevertheless the L-R deconvolution has restored numerous
elongated diffuse structures.  We have noted how the contrast of these
structures increases with successive iterations.  However, ultimately
the background noise becomes amplified into non-physical small-scale
structure, and so intermediate iterations produce the best compromise
for studying the details of the diffuse emission.  From the appearance
of the maps, it was judged that 32 iterations is optimal.
Subsequently, this field has been imaged by
\emph{Herschel}\footnote{http://herschel.esac.esa.int/Science\textunderscore
  Archive.shtml} \citep{barlow2010}, providing the desired ground truth
to verify the efficacy of the L-R deconvolution.  For comparison, the
250~\micron\ SPIRE image was smoothed with a 40\arcsec\ Gaussian, and
regridded to the 20\arcsec\ pixelization of the BLAST image.
Figure~\ref{fig:casa} shows this SPIRE image overlaid with contours from
the corresponding deconvolved 250~\micron\ BLAST map after 32
iterations.  We have again examined the higher-iteration images and
confirmed from this direct comparison that for the signal to noise of
this BLAST survey, 32 iterations is about optimal.

There is also good evidence for recovering diffuse structures in the
fields being studied in this paper.  The \MSX\ 8~\micron\ image of the
K3-50 region in Figure~\ref{fig:w58} shows emission (probably from PAHs)
along an extended shell-like structure, labelled ``arc."  Related
structure is also seen in the radio emission, suggesting a bubble and
surrounding PDR The deconvolved BLAST05 map reveals that the same
structure is also being traced by dust continuum emission, though the
effect of the map edge on the lower right is apparent too.  Furthermore,
there is a faint ridge of submillimeter dust emission extending to the
lower left of source K5 (K3-50C) which is traced by the CO contour.

\begin{figure}
\includegraphics[width=\linewidth]{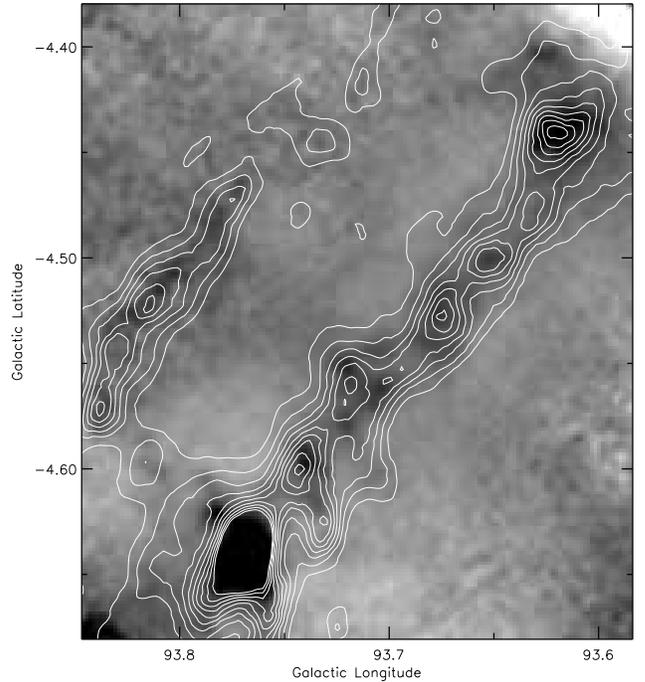}
\caption{Contours from the deconvolved BLAST 500~\micron\ image overlaid
  on the \SCUBA\ 850~\micron\ map of the ``Northern Streamer" in the
  IC~5146 region.  BLAST contours trace small-scale sub-structures in
  the \SCUBA\ map along the filament, again showing the ability of L-R
  deconvolution to restore morphological detail within diffuse
  emission.}
\label{fig:scuba}
\end{figure}

Within the IC~5146 field is an elongated filamentary molecular cloud
structure commonly known as the ``Northern Streamer."
Figure~\ref{fig:scuba} shows a \SCUBA\ 850~\micron\ dust map of a
targeted section in IC~5146 \citep{kramer2003,difrancesco}, overlaid
with contours from the 500~\micron\ deconvolved image.  This shows that
the L-R operation has preserved both large-scale structures and
smaller-scale fragments along the ridge.
A discussion of the astrophysics for this region is deferred to
\S~\ref{sec:ic5146}.

\section{ Science Results for K3-50}\label{sec:k350}

K3-50 is a well-known group of compact \HII\ regions, namely K3-50A--F,
within the star-forming complex W58
\citep{israel1976,howard1996,samal2010}.  The warm dust surrounding the
most prominent ionized region is bright in both far-infrared and
submillimeter emission; the correlation with the CGPS 21-cm radio
continuum emission is seen in Figure~\ref{fig:w58}.

The radio sources K3-50A and K3-50B are well centered inside the BLAST
clump K4 (spatial extent $\sim$ 1.7 pc), whereas K3-50C is within
K5. \cite{samal2010} further resolved K3-50C into two sources (C1 and
C2). Moreover, this young star-forming region contains embedded infrared
star clusters, namely K3-50D and K3-50B \citep{bica2003}.

In the overlap region between the BLAST survey and the FCRAO $^{12}$CO
molecular line survey there is good correspondence between peaks of
submillimeter and CO emission, which provides further insight into
global morphology and gas dynamics in the region.  The CO contours on
the lower right panel of Figure~\ref{fig:w58} show that K3, K4, and K5
are part of the same cloud (also see \S~\ref{sec:lr-on-srcs}).

\begin{deluxetable*}{lcccccccccc}
\tablewidth{0pt}
\small
\setlength{\tabcolsep}{0.02in}
\tablecaption{SED Best-Fit Parameters of BLAST Sources in the K3-50 Field}
\tablehead{
\colhead{ID}&
\colhead{BLAST Source} &
\colhead{$S_{250}$}&
\colhead{$S_{350}$}&
\colhead{$S_{500}$}&
\colhead{$S_{100}$}&
\colhead{$S_{60}$}&
\colhead{$T$} &
\colhead{$M$}& 
\colhead{$L$}&
\colhead{$L_{\rm bol}$} \\
\colhead{}&
\colhead{Name}&
\colhead{(Jy)}&
\colhead{(Jy)}&
\colhead{(Jy)}&
\colhead{(Jy)}&
\colhead{(Jy)}&
\colhead{(K)}&
\colhead{($10^2$ M$_{\odot}$)}&
\colhead{($10^4$ L$_{\odot}$)} &
\colhead{($10^4$ L$_{\odot}$)}\\
}
\startdata
K1 & J$200054+333055$                       & $105\pm10$          & $ 61 \pm5$     & $23 \pm3$      &    $70$\tablenotemark{$\rm a$}      & $14$\tablenotemark{$\rm a$}        & $20\pm1$    &  $28\pm5$ &   $1\pm 1$   & \nodata\\
K2 & J$200112+333802$                       & $ 96\pm9$         & $ 46\pm3$        & $13\pm1$     & $277$        & $175$       & $33\pm1$    &   $6\pm 1$       &  $3\pm 1$      &   \nodata \\
K3\tablenotemark{$\rm b$} & J$200137+333454$& $144 \pm20$      & $ 74\pm8$         & $20\pm8$     &  \nodata     & $ 308$      & $34\pm1$   &  $10\pm2$        &   $5\pm 1$      &   \nodata\\
K4 & J$200146+333214$                       & $ 1468 \pm19$     & $576\pm9$        & $245 \pm6$   & $11600$      & $12000$     & $44\pm4$    &  $68\pm14$       &   $152\pm54$     &  $192$\\
K5 & J$200154+333343$                       & $809 \pm25$       & $ 357\pm10$      & $184 \pm6$   & $3300$        & $ 3100$    & $35\pm2$    &   $55\pm14$      &  $36\pm4$      &   $39$ \\
\label{tab:k3-50}
\enddata
\tablenotetext{a}{Data used as upper limit in SED fit}
\tablenotetext{b}{IRAS 19597+3327A}
\end{deluxetable*}

% SEDs
In our preliminary work \citep{truch2008} we measured flux densities for
K3-50A and K3-50C (K4 and K5) by simultaneously fitting a model of two
2-dimensional Gaussians convolved with the corrupted BLAST PSF, where
sizes, positions, and amplitudes were free parameters of the fit.
In our present analysis, with the well-resolved deconvolved map, we fit
multiple Gaussians with a linear background model to extract flux
densities for individual sources.  Table~\ref{tab:k3-50} summarizes the
flux densities.

In order to achieve a broader multi-wavelength description of the SEDs,
we have extracted flux densities at 60 and 100~\micron\ from the IGA
maps.  For longer wavelength data points we have used the integrated
flux densities reported by \cite{thompson2006} at 450 and
850~\micron\ for K4.  Due to the restricted coverage of the
\SCUBA\ imaging, flux densities at 450 and 850~\micron\ are not
available for the other sources.  We have also used \MSX\ photometric
data in deriving the bolometric luminosity; however, we have not used
these data in the simple fit to the long-wavelength SED.

The SED-fitting model is a simplified single-temperature modified
blackbody with a fixed dust emissivity index of $\beta = 1.5.$ To obtain
the mass, we have used equation (2) of \cite{roycyg2010}, adopting the
same 250-\micron\ dust opacity $\kappa_{0}$ $=$ 10~cm$^2$~g$^{-1}$ and
dust to gas ratio $r$ $=$ 0.01.  We use the given distance of 8.7~kpc.
From the \HI\ absorption velocities and the CO emission, the sources are
in a weak ``Perseus arm'' feature (in $l-v$ or $b-v$ diagrams) beyond
the tangent point (at $\sim 5.5$~kpc), but not beyond the stronger
emission at $-60$~\kms.
Table~\ref{tab:k3-50} summarizes the physical properties, namely
temperature ($T$), luminosity ($L$), and mass ($M$) for the individual
BLAST sources.  The reported 1-$\sigma$ uncertainties of the physical
parameters are obtained from Monte-Carlo simulations as described in
\cite{chapin2008}.

\begin{figure}[t]
\includegraphics[scale=.5]{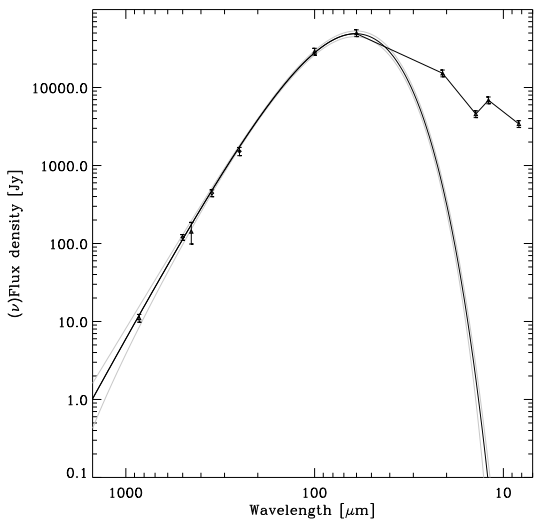}
\caption{SED for K3-50A (our K4). Flux density multiplied by
  $\nu/\nu_{250}$ plotted along the y axis. The solid line is the best
  fit to the modified blackbody with a fixed dust emissivity parameter
  of $\beta = 1.5.$ The grey lines represent an envelope of 68\%
  confidence level obtained from Monte-Carlo simulations. The best-fit
  parameters are given in Table ~\ref{tab:k3-50}.  Data at wavelengths
  lower than 60~\micron\ are not included in the fit but are used in
  estimating $L_{\rm bol}$.}
\label{fig:k3-50Ased}
\end{figure}

Figure~\ref{fig:k3-50Ased} shows an example of an SED fit, for K4.  A
best-fit temperature of 43~$\pm$~4~K and total mass of (7$\pm$
1)$\times$10$^3$ \msol\ are obtained from the parameters of the fit.
\cite{okamoto} have found a star cluster associated with K3-50A (the
dominant radio source inside K4, also an ultra-compact \HII\ region,
\citealp{kurtz1994}).  Our derived luminosity is
(1.5~$\pm$~0.5)$\times10^6$~\lsol.  This clump luminosity is equivalent
to a ZAMS spectral type of $\sim$O4 \citep{panagia1973}, which agrees
with \cite{samal2010}, who assigned a spectral type earlier than O5,
based on near-infrared measurements.  However the high luminosity
supports the possibility of multiple ionizing stars in the stellar
cluster.  In fact, for K4 it is important to note that we measure dust
emission whose energy comes collectively from K3-50A and K3-50B.

\begin{figure}[t]
\includegraphics[scale=.3]{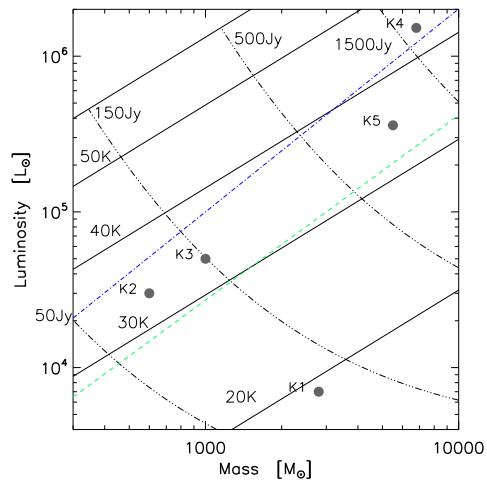}
\caption{BLAST compact sources in the K3-50 region in the
  \LM\ plane. Solid diagonal lines are loci of constant $T$ or $L/M$.
  Dot-dash curves “orthogonal” to these are for constant
  250~\micron\ flux density.  Dash and dot-dash lines denote sources
  powered by accretion and nuclear burning, respectively, as derived
  empirically in Fig.~9 of \cite{mol2008}.}
\label{fig:lm}
\end{figure}

The position of a source on the \LM\ plane provides a rough estimate of
its evolutionary stage \citep{mol2008}.  The relationship of the
evolutionary sequence to the underlying energetics powering the dust
emission is discussed in detail by \cite{roycyg2010}.
Figure~\ref{fig:lm} also confirms that K4 contains one or more zero age
main sequence stars, deriving luminosity mainly from nuclear burning,
and hot enough to cause copious ionization.

In a similar fashion, for K3-50C (K5) we obtained a temperature
35~$\pm$~2~K, which \cite{truch2008} underestimated due to confusion in
the maps combined with the uncertainties in the assumed model.
The energetics of K5 are primarily influenced by the two radio sources
inside the clump, namely K3-50C1 and K3-50C2 \citep{samal2010}. Compared
to K3-50A, the dust temperature is somewhat lower, suggesting that
K3-50C is deeply embedded inside the molecular cloud, is less energetic,
and is perhaps somewhat less evolved.  Integrating the SED for K3-50C,
we obtain a luminosity of (3.6 $\pm$ 0.4)$\times$10$^5$~\lsol\ which is
equivalent to $\sim$O6 on the ZAMS \citep{panagia1973}.  The clump mass
is (5.5~$\pm$~1.5)$\times$10$^3$~\msol, placing it in the appropriate
position in the \LM\ diagram.

K3 is located between the two radio sources K3-50E and K3-50F, to its
West and East, respectively (see~Fig.~\ref{fig:w58}).  Its luminosity is
(5~$\pm$~1)$\times$10$^4$~\lsol; \cite{samal2010} underestimated the
luminosity ($\sim$ 2.5$\times$10$^3$~\lsol) due to the absence of
FIR/submillimeter coverage.  The luminosity and mass of
(10~$\pm$~2)$\times$10$^2$~\msol\ position K3 in the \LM\ plane as a
``class I'' object, though with possibly some power coming from
accretion.  Its equivalent single-star ZAMS spectral type is $\sim$O8.5,
indicating the possibility of ionizing its surroundings.  No significant
radio continuum emission peak is detected either at 1420~MHz (see
Fig.~\ref{fig:w58}) or above the 5~mJy level at 1280~MHz
\citep{samal2010}.  Possibly there is radio self-absorption, or the
ionizing radiation is absorbed by dust in a dense
envelope. Alternatively, the luminosity may come from many slightly
lower mass stars, which, being cooler, would produce collectively less
ionizing radiation.

A general introduction of K1 and K2 was already given in
\S~\ref{sec:lr-on-srcs}. An accurate distance estimate for these sources
is not available. For our present analysis we have assumed a distance of
8.7~kpc, similar to the K3-50 region.  However, note that the assessment
of the evolutionary stages from the \LM\ diagram is not affected by the
distance uncertainty; the position on the \LM\ plane simply shifts
diagonally along a line of constant $L/M$ or temperature.
The K2 clump is associated with emission in the \IRAS\ 60 and
100~\micron\ bands and has extended \MSX\ 8~\micron\ emission as
well. It is relatively hot, with a temperature of 33~$\pm$~1~K. The
\LM\ plot suggests it is a less energetic version of K3, though the
\MSX\ morphology is distinctly different.
K1, projected at the outskirts of the K3-50 region, with no YSO or radio
counterpart, is the coldest BLAST clump in this region, at 19~$\pm$~1~K.
In the \LM\ diagram it is at an earlier stage of evolution, beginning to
be powered by accretion.

\section{Science results for IC~5146} \label{sec:ic5146}

The IC~5146 molecular cloud complex in Cygnus has been widely studied in
the optical, IR \citep{herbig2002}, submillimeter \citep{kramer2003} and
molecular lines \citep{dobashi1992}. Using infrared color excesses in
the \emph{JHK} bands, \cite{lada1994} produced an extinction map, i.e.,
the spatial distribution for the dust column density.  They also
surveyed in the molecular line emission of $^{13}$CO, C$^{18}$O, and CS.
\cite{herbig2008} provide a summary of the observational progress
achieved in this star-forming region.  \cite{harvey2008} have surveyed
the IC~5146 region with \emph{Spitzer} to study the properties of young
stellar objects.
Distance estimates range from 460 pc to 1.4~kpc (see
\citealp{harvey2008} and references therein). We have adopted 1~kpc from
\cite{dobashi1992}, which is close to the 950~pc used by
\cite{harvey2008}.  Proximity is an advantage for probing the workings
of star-forming regions.  At this distance, BLAST can resolve spatial
structures of about 0.3~pc.
Our BLAST05 observation (Fig.~\ref{fig:ic5146b}) targeted the
filamentary structure of the ``Northern Streamer," one of the densest
molecular clouds in IC~5146 (``cloud C'' of \citealp{dobashi1992}), and
reportedly the most massive ($\sim$2.2$\times10^3$~\msol).  The field
also contains the dark cloud L1030 and L1031 to the South-West.

\subsection{Compact Sources}\label{sec:5146csrcs}

% place in-line in text
\begin{deluxetable*}{lcrrrrrrrr}
\tablewidth{0pt}
\small
\setlength{\tabcolsep}{0.02in}
\tablecaption{Flux Densities of BLAST Sources in the IC~5146 Field}
\tablehead{
\colhead{ID}&
\colhead{BLAST Source}&
\colhead{$l$} &
\colhead{$b$} &
\colhead{$F_{250}$}&
\colhead{$F_{350}$}&
\colhead{$F_{500}$}&
\colhead{$F_{100}$}&
\colhead{$F_{60}$}&
\colhead{$F_{24}$}\\
\colhead{}&
\colhead{Name}&
\colhead{}&
\colhead{}&
\colhead{(Jy)}&
\colhead{(Jy)}&
\colhead{(Jy)}&
\colhead{(Jy)}&
\colhead{(Jy)}&
\colhead{(Jy)}\\
} %\tablenotemark{$\rm c $}
      %      ID                          name               l                 b         f250               f350               f500                      f100            f60              f24
\startdata
      IC1                           &  J$214558+471835$&  $ 93.4334$ &     $-4.6663$ &  $97.1\pm 9.0$ &   $75.1\pm7.3$     & $32.5\pm 2.1$&	        \nodata      &    \nodata     &	 \nodata\\
      IC2\tablenotemark{$\rm a $}   &  J$214508+473305$&  $ 93.4842$ &     $-4.3840$ &  $42.5\pm1.5$  &   $39.7\pm1.4$     & $22.4\pm 0.8$&	        $29.0\pm0.5$ &    $14.2\pm0.1$&	   $0.7$\\
      IC3                           &  J$214626+471757$&  $ 93.4844$ &     $-4.7141$ &  $60.1\pm4.7$  &   $42.3\pm2.3$     & $24.0\pm 1.0$&            \nodata      &    \nodata     &    $0.1$	\\
      IC4\tablenotemark{$\rm b $}   &  J$214452+474026$&  $ 93.5297$ &     $-4.2616$ &  $38.4\pm2.0$  &   $31.5\pm1.5$     & $19.9\pm 1.2$&            $24.3\pm0.3$ &   $13.3\pm0.01$&    $0.8$	\\
      IC5                           &  J$214534+473559$&  $ 93.5703$ &     $-4.3936$ &  $15.0\pm1.0$  &   $12.1\pm0.6$     & $8.7\pm0.4  $&             \nodata     &      \nodata   &    $0.1$\\         
      IC6\tablenotemark{$\rm c $}   &  J$214443+474635$&  $ 93.5769$ &     $-4.1670$ &  $45.1\pm3.4$  &   $33.5\pm1.7$     & $19.8\pm 1.2$&	        $10.1 \pm0.3$&   $2.3\pm0.1$  &	   $0.2$\\
      IC7                           &  J$214558+473536$&  $ 93.6132$ &     $-4.4397$ &  $69.1\pm6.5$  &   $47.7\pm2.3$     & $30.0\pm 1.3$&           $20.0\pm6.2$&   $1.7\pm0.1$  &	   $0.2$\\
      IC8                           &  J$214707+473242$&  $ 93.7369$ &     $-4.6054$ &  $23.0\pm2.6$  &   $20.6\pm1.5$     & $11.4\pm 1.5$ &             \nodata    &    \nodata     &      \nodata\\
      IC9\tablenotemark{$\rm d $}   &  J$214722+473221$&  $ 93.7659$ &     $-4.6372$ &  $186.6\pm3.4$ &   $121.8\pm2.2$    & $66.9\pm1.2$&             $97.5\pm10.0$&   $36.\pm2.2$  &	   $2.5$\\
      IC10                          &  J$214759+473643$&  $ 93.8913$ &     $ -4.6470$ & $45.4\pm2.7$  &   $30.8\pm1.3$     & $14.7\pm0.8$&	        $12.5\pm1.2$ &   $2.4\pm 0.2 $&	   $0.3$\\
\label{tab:ic5146}
\enddata
\tablenotetext{a}{IRAS 21433+4719} %ic2
\tablenotetext{b}{IRAS 21429+4726} %ic4
\tablenotetext{c}{IRAS 21428+4732} %ic6
\tablenotetext{d}{V1735 Cyg, Elias1-12} %ic9
\end{deluxetable*}

\begin{figure}
\includegraphics[width=\linewidth]{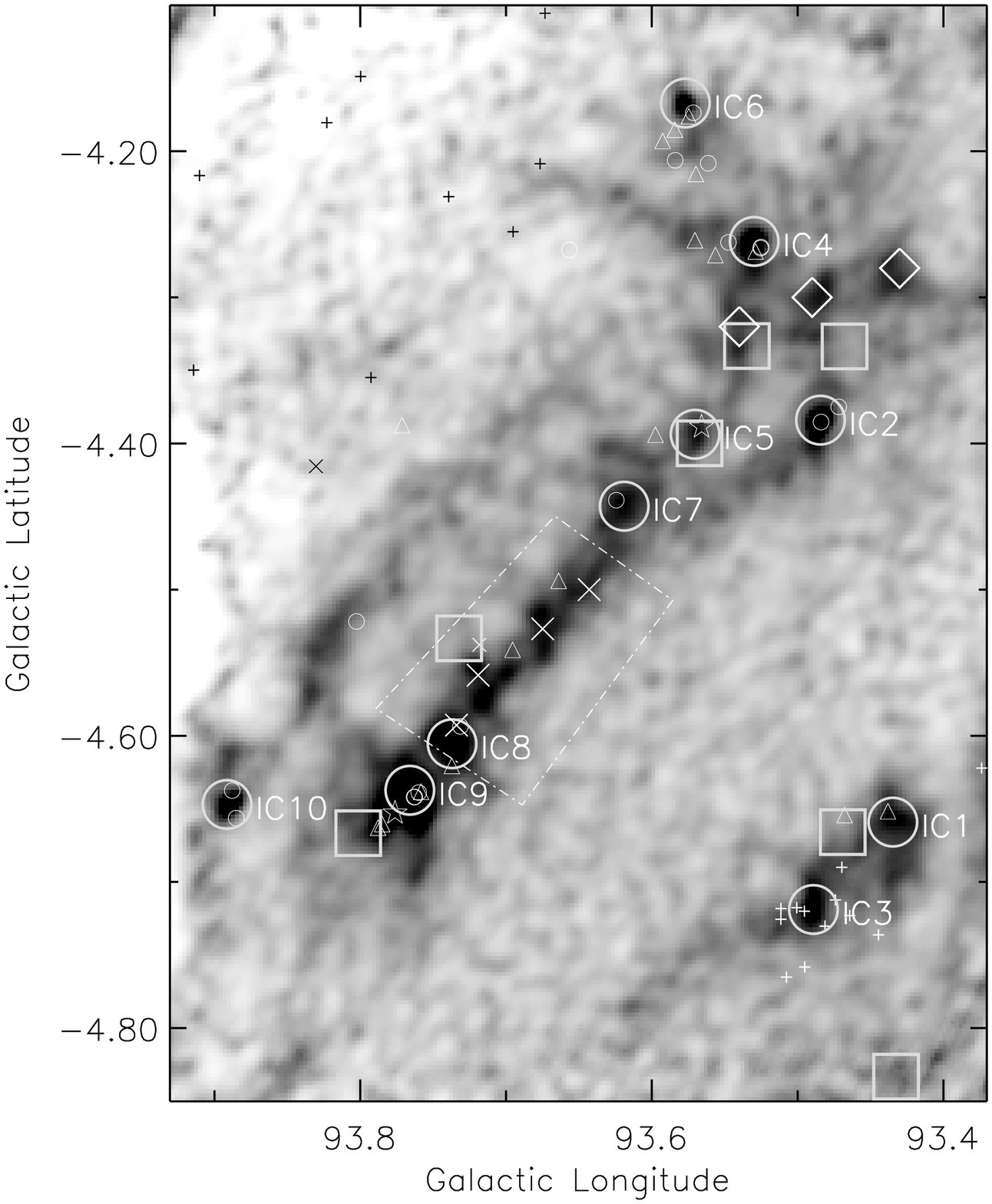}
\caption{BLAST map of IC~5146 at 250~\micron, showing relationships to
  other cataloged objects.
Circles of radius 1\arcmin\ locate the BLAST compact sources.
Large squares correspond to the peak positions of the
velocity-integrated $^{13}$CO ($J=1\rightarrow 0$) molecular line
emission map \citep{dobashi1992}.
YSOs from \citet{harvey2008} are shown by small circles, stars,
triangles, and crosses, to distinguish Class I, flat, II, and III
spectra, respectively; the plus symbols represent YSO candidates for
which there is no classification.
The dust temperature of the cold filament within the outlined dot-dashed
box is 11.7~$\pm$~0.4~K (\S~\ref{sec:streamer}).
Four cores characterized by \cite{kramer2003} are marked with large
crosses.
Some additional cold and starless clumps identified by BLAST are marked
by large diamonds.
}
\label{fig:ic5146b}
\end{figure}

Figure~\ref{fig:ic5146b} shows the locations of ten BLAST compact
sources in the IC~5146 region.  Six of these, with \IRAS\ counterparts,
were identified as protostellar candidates by \cite{dobashi1992}.  Among
these, Elias1-12 (our IC9) is a T-Tauri star \citep{elias1978}, also
known to be associated with molecular outflows \citep{levreault1983}.
Follow-up surveys with $^{12}$CO \citep{dobashi1993,dobashi2001} have
shown that all of these sources have molecular outflows, revealing the
early stages of star formation in progress inside the embedding clumps
seen by BLAST.
The squares in Figure~\ref{fig:ic5146b} correspond to the peak positions
of the velocity-integrated $^{13}$CO ($J=1\rightarrow 0$) molecular line
emission map. There is no direct correspondence between the BLAST dust
emission peaks and the $^{13}$CO peaks, but the BLAST compact sources
are associated with the CO clumps (see Fig.~6 of \citealp{dobashi1992}).

\cite{harvey2008} have detected 60 YSO candidates within the region
surveyed by BLAST. Using the slopes across \MIPS\ and \IRAC\ data they
classified 38 objects into Class~I, flat, II, and III. These relate to
progressively later stages of YSO evolution \citep{evans2009}. Their
positions and classifications are given in Figure~\ref{fig:ic5146b}.
Along the dense ``Northern Streamer" most of the sources belong to
Class~I and flat classifications, i.e., earlier stages of evolution.
Likewise, there are \emph{Spitzer} YSO counterparts to the ten BLAST
sources, these mostly being Class~I.  These YSOs within the more
extensive BLAST clumps (0.3~pc) are of relatively low power and so the
dust is not significantly internally heated and the BLAST dust
temperatures are all quite low.  For example, IC9 contains several YSOs
(in the 2MASS \emph{K}-band image there is reflection nebulosity). At
the present submillimeter sensitivity and resolution, not all YSOs have
BLAST counterparts.

We have characterized the physical properties of the identified BLAST
clumps by fitting SEDs as described in \S~\ref{sec:k350}. Compared to
the K3-50 region (dominated by warm dust emission), the dust temperature
of IC~5146 is relatively cold, which motivated us to constrain SEDs with
a spectral index of $\beta$ $=$ 2.0.

\begin{figure}
\includegraphics[scale=0.5]{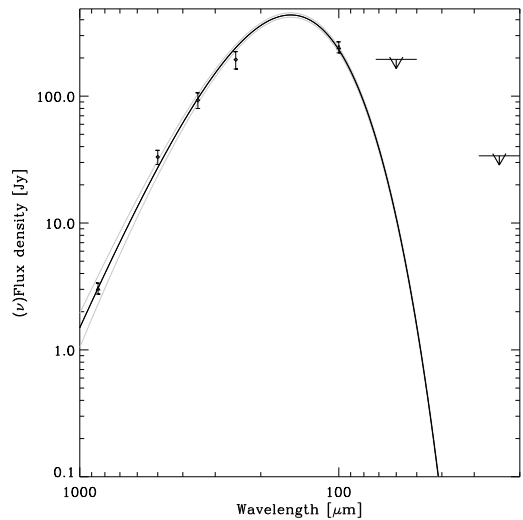}
\caption{SED as in Fig.~\ref{fig:k3-50Ased}, but for IC9 (Elias1-12) in
  the IC~5146 region fit with $\beta = 2$. Arrows for short-wavelength
  data represent 3-$\sigma$ upper limits above the measured flux
  density. The best-fit temperature for the cold envelope is 16~K.}
\label{fig:sedic}
\end{figure}

Where there are separable \IRAS\ sources, before performing photometry
on the BLAST sources, we have convolved the respective maps to the
2\arcmin\ native beam resolution of the IGA 100~\micron\ images
\citep{cao1997}. Otherwise we used the better BLAST resolution.
Table~\ref{tab:ic5146} summarizes the flux densities.  All of these
sources have 250~\micron\ flux densities greater than at 100~\micron,
indicating a relatively cold dust temperature.
Constraining the SEDs with the longer wavelength data points is
important, but unfortunately \SCUBA\ has observed only a small part of
the IC~5146 region (Fig.~\ref{fig:scuba}).  This includes IC7, IC8 and
IC9, for which we obtained integrated 850-\micron\ flux densities of
10.6, 1.5, and 5.6 Jy, respectively.
For the cool sources here we have used only the 100~\micron\ flux
density in the SED fits, while the 60 and 24~\micron\ flux densities
provide upper limits, as discussed in \cite{chapin2008} and
\cite{truch2008}.  As an example we show the SED for the brightest
clump, IC9, in Figure~\ref{fig:sedic}.

% inline table
\begin{deluxetable}{cccc}
\tablewidth{0pt}
\small
\setlength{\tabcolsep}{0.02in}
\tablecaption{SED Best-Fit Parameters of BLAST Sources in the IC~5146 Field}
\tablehead{
\colhead{ID}&
\colhead{$T$}&
\colhead{$M_{\rm tot}$}&
\colhead{$L$}\\
\colhead{}&
\colhead{(K)}&
\colhead{(M$_{\odot}$)}&
\colhead{(L$_{\odot}$)}\\
}
\startdata
      IC1  &   $12.7\pm 1.2 $  & $171\pm50$ & $48\pm 16$ \\
      IC2  &  $16.3 \pm 0.6 $  & $35\pm8$ & $43\pm 4$\\
      IC3  &  $14.9\pm 0.1 $   & $64\pm 6$&    $47\pm5$\\
      IC4  &  $16.1\pm 0.7 $  & $33\pm9$&    $38\pm4$\\
      IC5  &  $14.9\pm 0.1 $  & $15\pm2$&   $11\pm2$\\
      IC6  &  $13.6 \pm 0.3 $  & $73\pm11$ &    $31\pm 2$\\
      IC7  &  $14.0 \pm 0.3 $  & $106\pm 10$ &    $53\pm3$\\
      IC8  &  $15.0\pm 0.1$     & $25\pm3$   &    $19\pm 2$\\
      IC9  &  $15.7 \pm 0.3 $  & $176\pm18$ &    $171\pm12$\\
      IC10  & $14.3 \pm 0.3 $  & $55\pm7$ &    $31\pm2$\\
\label{tab:table3}
\enddata
\end{deluxetable}

Table~\ref{tab:table3} summarizes the physical quantities $T$, $M$, and
$L$ obtained from the best-fit parameters of the SEDs.  In this region
of IC~5146 the luminosities of the identified BLAST sources range from
11 to 170~\lsol\ and the masses from 15 to 180~\msol.  The temperatures
range from 13 to 16~K, with an average about 14.8~K.

When a clump is more massive than the Bonner-Ebert critical mass, then
gravity overpowers the internal (thermal, magnetic, and turbulent)
supporting pressure and it undergoes collapse.  Our BLAST mass estimates
appear to be above the critical mass (using the simple prescription
described in \citealp{kerton2001}), suggesting that the compact sources
are gravitationally unstable.  Indeed, there are already some YSOs that
are generating outflows.
\cite{dobashi2001} showed how physical processes relating to mass,
momentum, and energy carried away by the outflows have direct
consequences for the stability of the parent cloud and as well influence
the evolutionary dynamics of the embedded stars.  These outflow
parameters have a correspondence with the bolometric
luminosity of the accreting source \citep{dobashi2001}.  Certainly
feedback processes have already begun in this molecular cloud, and
\cite{dobashi1992} have suggested that outflows have played an important
role in supporting the parent cloud from collapsing.

\subsection{Cold Central Filament}\label{sec:streamer}

Apart from the above-mentioned compact sources, BLAST observes abundant
substructure in the Northern Streamer region.  The infrared extinction
map by \cite{lada1994} has also revealed a distribution of high column
density clumps throughout this filament.  There is a good correlation of
$A_{\rm V}$ with the $^{12}$CO and $^{13}$CO molecular line emission
maps by \cite{dobashi1992}.  Not unexpectedly, our dust emission map
also has a strikingly tight correlation with $A_{\rm V}$ (and CO which
might not be a perfect tracer of column density at large spatial
density), and with 850~\micron\ emission, as shown in
Figure~\ref{fig:scuba}.

With earlier \SCUBA\ observations \cite{kramer2003} studied a narrow
14\arcmin\ by 2\farcm5 region extending from IC7 (only at the boundary
of their map) to IC8 (included).  From the \SCUBA\ colors they found
dust temperatures ranging up to 20~K at the outskirts and between column
density peaks and down to 10~K in ``cores'' (coincident with peaks in
$A_{\rm V}$), embedded condensations that effectively shield the
interstellar radiation from which the dust derives its power in the
absence of a YSO.  Note that along the central filament studied there
are no embedded YSOs (Fig.~\ref{fig:ic5146b}), except at the two ends.

\begin{figure}
\includegraphics[scale=.4]{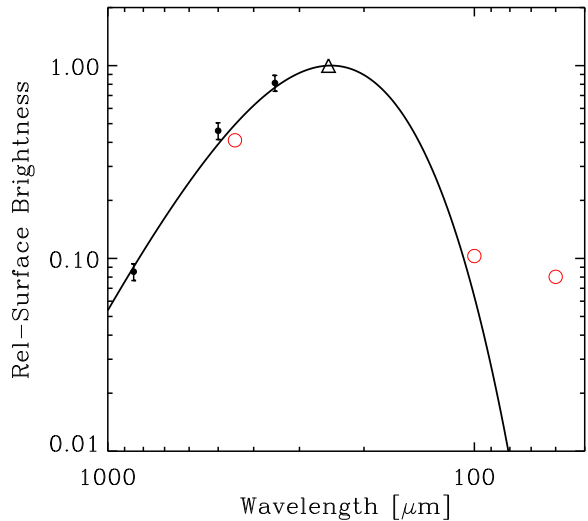}
\caption{Relative SED obtained from pixel-pixel image correlation with
  respect to 250~\micron\ emission for the area within the dot-dashed
  box marked in Fig.~\ref{fig:ic5146b}.  With the dust emissivity index
  $\beta$ fixed at 2.0, the best-fit temperature is 11.7 $\pm$ 0.4~K.
  Data points represented by circles are not actually used in the fit,
  while the triangle shows the value of unity at the normalizing
  250~\micron\ point.}
\label{fig:relsed}
\end{figure}

We have estimated a global dust temperature for the portion of the
streamer within the box shown in Figure~\ref{fig:ic5146b}, by fitting a
relative SED to the pixel-pixel correlation slopes of the image data
with respect to 250~\micron.  We have used the 850~\micron\ \SCUBA\ map
to constrain the spectrum at long wavelengths.  The filament is not
prominent in \IRAS\ bands because the dust is cold.
Figure~\ref{fig:relsed} shows the SED.  With a dust emissivity index
$\beta$ of 2.0 the best-fit temperature is 11.7~$\pm$~0.4~K, which
agrees well with \cite{kramer2003}.

\subsection{Cold Starless Clumps}\label{sec:cold}

Because the BLAST bands are sensitive to colder dust emission we can
detect cold clumps early in their pre-stellar evolution, when their
primary energy source is the external radiation field (``stage E" in
\citealp{roycyg2010}).  Good examples confirmed by BLAST are the four
``cores'' identified by \cite{kramer2003} and marked by large crosses in
Figure~\ref{fig:ic5146b}.  The core numbered 2 by \cite{kramer2003} is
the warmest (18.2~K). This is a complex region near IC8 with large
temperature gradients influenced by a YSO visible even to \IRAS\ at
short wavelengths. However the other three cores are cold and apparently
starless.

With the larger spatial coverage of the BLAST survey we have identified
other cold clumps that have no signs of star formation yet.  Some
prominent examples are marked by large diamonds in
Figure~\ref{fig:ic5146b}, at the positions of the peaks in the
500~\micron\ map.  Comparing with the visual extinction map of
\cite{lada1994}, we find that these BLAST clumps G93.43$-$4.28,
G93.48$-$4.30, and G93.54$-$4.33 have $A_{\rm V}$ about 10, 20, and 6
mag, respectively.

The morphology in this field is complex and variable with wavelength,
because of the different stages of evolution occurring in close
proximity.  For example, the protostar IC4 occurs at the end of an
elongated condensation that has a cold extension in the direction of
G93.54$-$4.33. Studies of this region will benefit from the improved
resolution of \emph{Herschel}.  With map zero points available from
Planck, it will be interesting to map the temperature structure in
detail, as has been accomplished in other fields
\citep{bernard2010,juvela2010}.

\section{Conclusion}

The Lucy-Richardson deconvolution algorithm has been applied
successfully to improve BLAST05 images obtained with a corrupted (but
known) PSF.  This deconvolution has enhanced the raw BLAST map of
$\sim$3\farcm3 resolution to $\sim$1\arcmin, near the anticipated
diffraction limit.  This improves the detectability of faint sources;
diffuse structures are revealed in finer detail as well.
We have checked via simulations the robustness of applying the L-R
scheme, especially the aspects of conserving flux, reliability of
restoring intrinsic sizes, and performance in resolving sources in a
crowded field.
For the actual deconvolved maps of BLAST05 survey fields we have
provided further ground truth for the improved detail in crowded fields
and for diffuse structures, by comparing with available multi-wavelength
high resolution images of dust and other tracers.

We have presented science results enabled by the deconvolved maps of two
star-forming regions covered by BLAST05, namely K3-50 and IC~5146.  We
were able to resolve three crowded sources in K3-50, namely K3-50A,
K3-50C, and IRASB, and have also characterized another two sources which
had previously remained undetected.

The deconvolved maps of IC~5146 have further shown the richness of the
field, consisting of both large scale diffuse structures and compact
sources, with a considerable range of dust temperature because of the
different stages of evolution.  The compact BLAST sources characterized
here all have associated YSOs. However, these are not so powerful as to
heat up the entire dust clump measured by BLAST, so that $L/M$ $<$
1~\lsol/\msol\ and the effective dust temperature is rather low; star
formation is just getting under way.  In fact elsewhere there are high
column density structures that are starless, like within the central
filament in the Northern Streamer, for which we have obtained an average
dust temperature of 11.7~K.  We have found further examples in the wider
BLAST field surveyed.

\acknowledgments
The BLAST collaboration acknowledges the support of NASA through grant
numbers NAG5-12785, NAG5-13301, and NNGO-6GI11G, the Canadian Space
Agency (CSA), the UK Particle Physics \& Astronomy Research Council
(PPARC), and Canada's Natural Sciences and Engineering Research
Council (NSERC). We would also like to thank the Columbia Scientific
Balloon Facility (CSBF) staff for their outstanding work.

\bibliographystyle{apj}
%\bibliography{lr,myrefn}

\end{document}